\documentclass[prd,a4paper]{revtex4}
\usepackage{graphicx}
\usepackage{dcolumn}
\usepackage{bm}
\usepackage{epsfig}
\newcommand{\be}{\begin{equation}}
\newcommand{\ee}{\end{equation}}
\newcommand{\bea}{\begin{eqnarray}}
\newcommand{\eea}{\end{eqnarray}}
\newcommand{\bean}{\begin{eqnarray*}}
\newcommand{\eean}{\end{eqnarray*}}

\newcommand{\gapproxeq}{\lower
.7ex\hbox{$\;\stackrel{\textstyle >}{\sim}\;$}}
\newcommand{\lapproxeq}{\lower
.7ex\hbox{$\;\stackrel{\textstyle <}{\sim}\;$}}

\begin{document}

\bibliographystyle{unsrt}

\title{\bf Study of isospin violating $\phi$ excitation in $e^+e^- \to \omega\pi^0$}

\author{Gang Li$^{1,3}$,  Yuan-Jiang Zhang$^{1,3}$ and  Qiang Zhao$^{1,2,4}$}

\affiliation{1) Institute of High Energy Physics, Chinese Academy of
Sciences, Beijing 100049, P.R. China}

\affiliation{2) Department of Physics, University of Surrey,
Guildford, GU2 7XH, United Kingdom}

\affiliation{3) Graduate University of Chinese Academy of Sciences,
Beijing 100049, P.R. China}

\affiliation{4) Theoretical Physics Center for Science Facilities,
Chinese Academy of Sciences, Beijing 100049, P.R. China}

\date{\today}

\begin{abstract}

We study the reaction $e^+ e^-\to \omega\pi^0$ in the vicinity of
$\phi$ mass region. The isospin-violating $\phi$ excitation is
accounted for by two major mechanisms. One is electromagnetic (EM)
transition and the other is strong isospin violations. For the
latter, we consider contributions from the intermediate hadronic
meson loops and $\phi$-$\rho^0$ mixing as the major mechanisms via
the $t$ and $s$-channel transitions, respectively. By fitting the
recent KLOE data, we succeed in constraining the model parameters
and extracting the $\phi\to\omega\pi^0$ branching ratio. It shows
that the branching ratio is sensitive to the $\phi$ excitation line
shape and background contributions. Some crucial insights into the
correlation between isospin violation and Okubo-Zweig-Iizuka (OZI)
rule evading transitions are also learned.

\end{abstract}

\maketitle

 PACS numbers: 12.40.Vv, 13.20.Gd, 13.25.-k

\vspace{1cm}

\section{Introduction}
The $e^+e^-$ annihilation at low energies, below 3 GeV, is a major
source of information on the properties of light vector mesons
$(\rho,\omega,\phi)$ and their excited states. Recently, the cross
sections of $e^+e^-\to\omega\pi^0$ are measured by KLOE
collaboration in the vicinity of
$\phi$-resonance~\cite{Ambrosino:2007wf,Ambrosino:2008gb}, where a
dip appears at the $\phi$ mass. It greatly improves the experimental
status for the isospin-violating $\phi$ meson production in the $e^+
e^-$ annihilation. Meanwhile, the significant difference in the
extraction of the $\phi\to\omega\pi^0$ branching ratio in
Refs.~\cite{Ambrosino:2007wf,Ambrosino:2008gb} suggests how
difficult to measure such a tiny effect in $e^+ e^-$ annihilations.
This transition is correlated with the Okubo-Zweig-Iizuka (OZI) rule
violation due to the nearly-ideal flavor mixing between $\omega$ and
$\phi$. Since the reaction $e^+ e^-\to \omega\pi^0$ is dominated by
the $I=1$ transition matrix element, the interferences due to the
$\phi$ production will provide an opportunity for probing the
isospin-violating mechanism and its correlation with the OZI rule.

In Ref.~\cite{Li:2007au}, the isospin-violating decay $\phi\to
\omega\pi^0$ has been studied in an effective Lagrangian approach.
Two mechanisms are quantifies as the major sources for isospin
violations. One is the electromagnetic (EM) transition where the
$s\bar{s}$ annihilates into a virtual photon and then decays into
$\omega$ and $\pi^0$. At leading order, the transition can be
accounted for by virtual photon exchange between isoscalar and
isovector mesons in vector meson dominance (VMD) model. The isospin
conserved vertices can be well constrained by experimental data for
$\omega\to\gamma\pi^0$, $\omega\to e^+ e^-\pi^0$, $\phi\to e^+
e^-\pi^0$, and $\phi\to\rho\pi +3\pi$~\cite{pdg2006}. The other is
the strong isospin violation originated from the $u$-$d$ quark mass
difference due to chiral symmetry breaking of QCD
vacuum~\cite{Miller:1990iz}. A dynamic process to recognize this is
via the OZI-rule-evading intermediate meson loop
transitions~\cite{Li:2007au}.

In this work we study the $\phi$ meson excitation in $e^+ e^-\to
\omega\pi^0$ in the effective Lagrangian approach in order to
further improve our understanding of the strong isospin violation
mechanism. In contrast with $\phi\to \omega\pi^0$~\cite{Li:2007au},
more transition mechanisms will be involved in and contribute to
$e^+ e^-\to \omega\pi^0$. Detailed analysis of those background
contributions will be carried out. Compared with other approaches
via $\pi$-$\eta$ mixing or empirical $\phi$-$\omega$-$\rho^0$
mixing~\cite{bramon-81,donoghue,ametller,coon-86,coon-87,genz}, our
model is based on explicit experimental facts and constructed in a
self-consistent way at the tree level of the effective Lagrangian
approach.

As follows, we first analyze the mechanism for $I=1$ transitions in
the VMD model. We then include the interferences from the $I=0$
$\phi$ production based on the improved prescription of
Ref.~\cite{Li:2007au}. Numerical results are presented in Sec. III
and a summary is given in Sec. IV.

\section{The model}

\subsection{$I=1$ transitions in VMD model}

The $I=1$ EM transition amplitudes are described by isovector meson
fields. In the energy region around $M_\phi$ mass, the dominant
contributions are due to the low-lying vector mesons, such as
$\rho^0$, $\rho^\prime(1450)$, etc. In the VMD
model~\cite{Bauer:1975bw}, we can decompose the virtual photon by
these mesons, e.g. $\rho$ and $\rho^\prime$, as shown in Fig.
\ref{fig-3}(a) and (b). The transition amplitudes corresponding to
Fig.~\ref{fig-3}(a) and (b) are as follows: \bea M_{fi}^a &= &{\bar
v}^{(s^\prime)}(p_e^\prime)(-ie\gamma^\beta)u^s(p_e) \frac {-1}
{s(s-M_\rho^2 + i\Gamma_\rho(s) \sqrt{s})}\frac {eM_\rho^2}
{f_\rho}\frac {g_{\omega \rho\pi}} {M_\omega}
\varepsilon_{\alpha\beta\mu\nu}p_\rho^\alpha
p_\omega^\mu\varepsilon_\omega^\nu ,\\
M_{fi}^b &= &{\bar
v}^{(s^\prime)}(p_e^\prime)(-ie\gamma^\beta)u^s(p_e) \frac {-1}
{s(s-M_{\rho^\prime}^2 + i\Gamma_{\rho^\prime}(s) \sqrt{s})}\frac
{eM_{\rho^\prime}^2} {f_{\rho^\prime}}\frac {g_{\omega
{\rho^\prime}\pi}} {M_{\rho^\prime}}
\varepsilon_{\alpha\beta\mu\nu}p_{\rho^\prime}^\alpha
p_\omega^\mu\varepsilon_\omega^\nu ,
 \eea
 where $eM_V^2/f_V$ is a
direct photon-vector-meson coupling in Feynman diagram language and
$g_{\omega \rho\pi}$ and $g_{\omega \rho^\prime\pi}$ are the $VVP$
strong coupling constants.  $\Gamma_\rho(s)$  and
$\Gamma_{\rho^\prime}(s)$ are energy-dependent total widths for the
intermediate $\rho$ and $\rho^\prime$, respectively. They are
described by the two major decay modes, $\pi^+\pi^-$ and
$\omega\pi^0$~\cite{Akhmetshin:2003ag}: \bea \Gamma_\rho(s) &=&
\Gamma_\rho(M_\rho^2)\frac {M_\rho^2} {s} \left(\frac {p_\pi(s)}
{p_\pi(M_\rho^2)}\right)^3 + \frac
{g_{\omega\rho\pi}^2}{12\pi M_\omega^2} p_\omega^3(s), \label{rho-decay-width}\\
\Gamma_{\rho^\prime}(s) &= & \Gamma_{\rho^\prime}
(M_{\rho^\prime}^2) \left[{\cal B}_{\rho^\prime\to\omega\pi^0}
\left(\frac {p_\omega(s)} {p_\omega(M_{\rho^\prime}^2)}\right)^3 +
(1-{\cal B}_{\rho^\prime\to\omega\pi^0}) \frac {M_{\rho^\prime}^2}
{s} \left(\frac {p_\pi(s)}
{p_\pi(M_{\rho^\prime}^2)}\right)^3\right],\label{rhoprime-width}
\eea where the coupling constant $g_{\omega\rho\pi}/M_\omega$ is
taken to be 17 GeV$^{-1}$, which is the fitted value using three
different fitting schemes in Ref.~\cite{Akhmetshin:2003ag}. This
value is consistent with the experimental values derived from
$\omega\to\pi^0\gamma$, $\rho\to\pi^0\gamma$ and
$\omega\to\rho\pi\to \pi^+\pi^-\pi^0$ decays, and theoretical
estimates based on the QCD sum rules
~\cite{Eletsky:1982py,Narison:1982yc,Lublinsky:1996yf}.

It should be noted that the second term of
Eq.~(\ref{rho-decay-width}) does not contribute below the
$\omega\pi^0$ threshold. For $\rho^\prime(1450)$, its total and
partial widths still have large undertainties~\cite{pdg2006}. Its
best known decay channel is $\omega\pi$. Therefore, we parameterize
its energy-dependent total width into two parts, i.e.
$\rho^\prime\to\omega\pi$ and $\rho^\prime$ decay into other final
states. ${\cal B}_{\rho^\prime\to\omega\pi^0}$ is the branching
ratio of $\rho^\prime\to \omega\pi^0$. In the consequent
calculation, we will fix $\Gamma_\rho(M_\rho^2) = 149.6$ MeV and
take $\Gamma_{\rho^\prime}$ as a parameter due to its large
uncertainties.

In the above treatment, we assume that the dominant contributions to
the $I=1$ component are from $\rho$ and $\rho^\prime$ in the energy
around $M_\phi$ mass. Contributions from higher $I=1$ states should
be largely suppressed since their mass are larger than the
virtuality of the photon. So, the $I=1$ total transition amplitude
can be expressed as:
 \be\label{I1-em} M_{fi}^{I=1} = M_{fi}^a +
e^{i\delta_1}M_{fi}^b,
 \ee
 where a possible relative phase between
these two amplitudes is included and will be determined by
experimental data. We mention in advance that the dominant $I=1$
background requires $\delta_1=180^\circ$ as a constructive phase
between $\rho$ and $\rho^\prime$ terms in Eq. (\ref{I1-em}).

\subsection{$I=0$ transitions}

As studied in Ref.~\cite{Li:2007au}, the excitation of $\phi$ meson
can couple to $\omega\pi^0$ via two isospin violation processes. One
is EM transitions through $\phi$ meson annihilation. The other is
OZI-evading intermediate meson loops. We separately investigate
these two processes in this section. We also consider background
amplitudes from the off-shell $\omega$ meson excitations.

\subsubsection{$I=0$ EM transitions in VMD model}

\begin{figure}
\begin{center}
\includegraphics[scale=0.8]{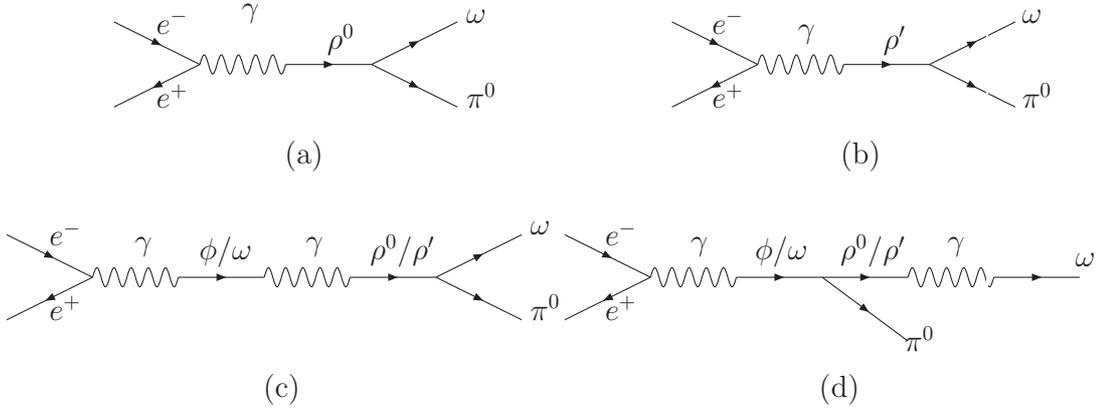}
\caption{The schematic diagrams for $I=1$ [(a) and (b)] and $I=0$ EM
transitions [(c) and (d)].}\protect\label{fig-3}
\end{center}
\end{figure}

The excitations of the $\phi$  and $\omega$ meson contribute to the
signal and background, respectively. We decompose the virtual photon
by a sum of vector mesons as shown in Fig. \ref{fig-3}(c) and (d).
The corresponding amplitudes have the following expressions:
 \bea\label{I0-c} \nonumber{\cal M}_{fi}^{c-V}
&=& \sum_{V_1,V_2} {\bar
v}^{(s^\prime)}(p_e^\prime)(-ie\gamma^\nu)u^s(p_e) \frac {1}
{s^2(s-M_{V_1}^2+i\Gamma_{V_1}(s) \sqrt{s})(s-M_{V_2}^2+i\Gamma_{V_2}(s)\sqrt{s})} \\
&\times &(\frac {eM_{V_1}^2} {f_{V_1}})^2 \frac {eM_{V_2}^2}
{f_{V_2}} \frac {g_{\omega V_2\pi}} {M_\omega}
\varepsilon_{\alpha\beta\mu\nu}{p_\omega^\alpha
\varepsilon_\omega^\beta p_\pi^\mu },
 \eea
  and
  \bea\label{I0-d}
\nonumber{\cal M}_{fi}^{d-V} &=& \sum_{V_1, V_2} {\bar
v}^{(s^\prime)}(p_e^\prime)(-ie\gamma^\nu)u^s(p_e) \frac {1}
{s(s-M_{V_1}^2+i\Gamma_{V_1}(s) \sqrt{s})(p_\omega^2-M_{V_2}^2+i\Gamma_{V_2}(M_\omega^2)M_\omega)M_\omega^2} \\
&\times &\frac {eM_{V_1}^2} {f_{V_1}} \frac {eM_{V_2}^2} {f_{V_2}}
\frac {eM_\omega^2} {f_\omega}\frac {g_{V_1 V_2\pi}} {M_\phi}
\varepsilon_{\alpha\beta\mu\nu}{p_\omega^\alpha
\varepsilon_\omega^\beta p_\pi^\mu }, \eea where $V_1$ denotes
$\phi$ or $\omega$ mesons, while $V_2$ denotes $\rho$ or
$\rho^\prime$ mesons; $g_{V_1V_2\pi}$ are the $VVP$ strong coupling
constants and $\Gamma_V(s)$ is the total width of the intermediate
vector mesons. The $V\gamma$ coupling values can be found in
Ref.~\cite{Li:2007ky}. The width $\Gamma_\phi(s)$ is described by
its major decay modes, $K^+K^-$, $K_L^0 K_S^0$ and
$\rho\pi+\pi^+\pi^-\pi^0$:
\bea \nonumber \Gamma_{\phi}(s) &= & \Gamma_\phi (M_\phi^2)
\left[{\cal B}_{\phi\to K^+K^-}  \frac {M_\phi^2} {s}\left(\frac
{p_{K^+}(s)} {p_{K^-}(M_\phi^2)}\right)^3 + {\cal B}_{\phi\to
K_L^0K_S^0} \frac {M_\phi^2} {s}\left(\frac {p_{K_L^0}(s)}
{p_{K_L^0}(M_\phi^2)}\right)^3\right.
 \\ &+&\left.(1-{\cal B}_{\phi\to K^+K^-}-{\cal B}_{\phi\to K_L^0K_S^0}) \left(\frac {p_\rho(s)} {p_\rho(M_\phi^2)}\right)^3
 \right],\label{phi-width}
\eea where ${\cal B}_{\phi\to K^+K^-}$, ${\cal B}_{\phi\to
K_L^0K_S^0}$ are the branching ratios of $\phi\to K^+K^-$, $\phi\to
K_L^0K_S^0$, respectively. Coupling $g_{\phi\rho^0\pi^0}=0.68$ is
extracted~\cite{Li:2007au} from the KLOE result for $\phi\to
\rho\pi+\pi^+\pi^-\pi^0$~\cite{kloe-phi-rho-pi}.

The relative sign between $\rho$ and $\rho^\prime$ intermediate
vector mesons is the same as $I=1$ transitions, and there is a
relative sign $\delta_2$ between the processes Fig. \ref{fig-3} (c)
and (d), i.e.
 \bea {\cal M}_{fi}^{EM-I=0}=({\cal
M}_{fi}^{c-\rho}+e^{i\delta_2} {\cal
M}_{fi}^{d-\rho})+e^{i\delta_1}( {\cal M}_{fi}^{c-\rho^\prime}+
e^{i\delta_2} {\cal M}_{fi}^{d-\rho^\prime}) \ .
 \eea
In the calculation we fix $\delta_2=0^\circ$ as given by the
effective Lagrangians, while $\delta_1=180^\circ$ is required as a
constructive phase between the transitions mediated by $\rho$ and
$\rho^\prime$ in Fig. \ref{fig-3} (c) and (d).

\subsubsection {$I=0$ transitions via intermediate meson loops}

\begin{figure}
\begin{center}
\includegraphics[scale=0.8]{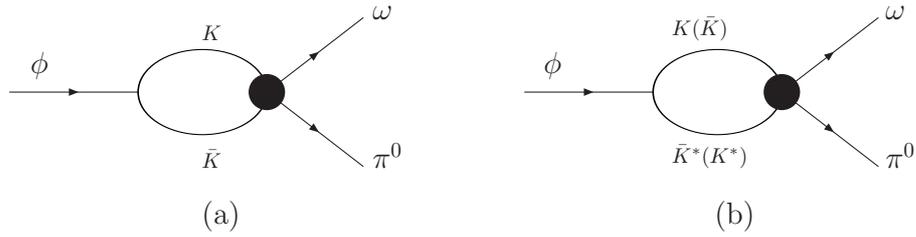}
\caption{The schematic diagrams for $\phi\to\omega\pi^0$ via the
OZI-rule-evading intermediate meson loop
transitions.}\protect\label{fig-1}
\end{center}
\end{figure}

\begin{figure}
\begin{center}
\includegraphics[scale=0.8]{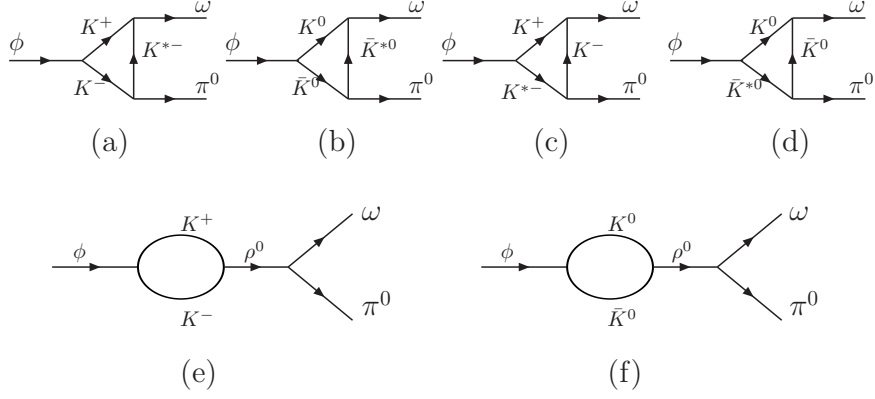}
\caption{The transition mechanisms for $\phi\to\omega\pi^0$ via
$t$-channel [(a)-(d)] and $s$-channel transitions [(e) and
(f)].}\protect\label{fig-2}
\end{center}
\end{figure}

The strong isospin violation via the OZI-rule-evading intermediate
meson loop transitions can be established by two experimental
observations. One is the obvious mass difference between the charged
and neutral kaon as a consequence of the $u$-$d$ quark-mass
difference. The other is the precise measurement of the $\phi\to
K^+K^-$ and $K_L K_S$ branching ratios, where the effective
couplings of $\phi K^+K^-$ and $\phi K^0\bar{K^0}$ turn to be
slightly different. Therefore, we can describe the strong isospin
violation as a dynamic outcome of Fig.~\ref{fig-1} for which the
detailed mechanisms can be recognized by transitions via
Fig.~\ref{fig-2}. In particular, transitions of
Fig.~\ref{fig-2}(a)-(d) can be classified as $t$-channel processes,
while (e) and (f) are $s$-channel ones.

In both kinds of transitions, the $u$-$d$ quark-mass difference
leads to the mass difference between the charged and neutral kaon,
which then results in nonvanishing $I=1$ cancellations between the
charged and neutral kaon loops as a source of the strong isospin
violations in $\phi\to\omega\pi^0$. Contributions from
$K\bar{K^*}(K)$ and $K\bar{K^*}(K^*)$ loops are also considered in
the $t$-channel where the effective couplings for the charged and
neutral particles are treated in the SU(3) symmetry limit as a
leading approximation. The $s$-channel transitions of
Fig.~\ref{fig-2}(e) and (f) contribute to the mixing of the $\phi$
and $\rho^0$, which were not considered in Ref.~\cite{Li:2007au}.
However, we state in advance that their contributions are not as
large as the $t$-channel transitions. In this work, in order to keep
this effective Lagrangian approach self-consistent, we explicitly
include the $s$-channel $K\bar{K}$ loop in the calculation. The
contributions from the $K\bar{K^*}+c.c.$ loops in the $s$ channel
are found even smaller since their mass threshold is higher than the
$\phi$ mass. Therefore, we do not consider them in the later
discussions.

\begin{figure}
\begin{center}
\includegraphics[scale=0.8]{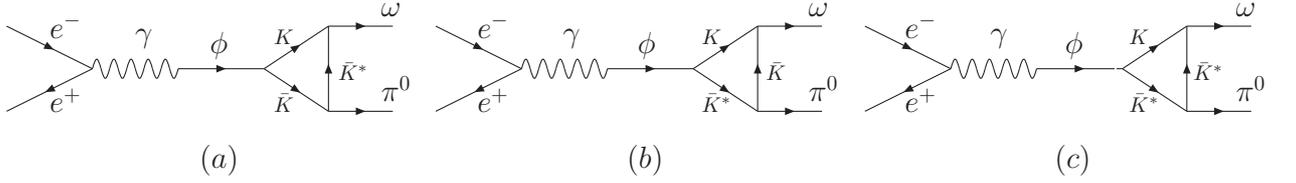}
\caption{ The schematic diagrams of $I=0$ transitions for
$e^+e^-\to\omega\pi^0$ via $KK(K^\ast)$, $KK^\ast(K)$ and
$KK^\ast(K^\ast)$ intermediate meson loops.}\protect\label{fig-4}
\end{center}
\end{figure}

In principle, we should include all the possible intermediate meson
exchange loops in the calculation. In reality, the break-down of the
local quark-hadron duality allows us to pick up the leading
contributions as a reasonable
approximation~\cite{Lipkin:1986bi,Lipkin:1986av}. As explained in
Ref.~\cite{Li:2007au}, the leading contributions arise from
$K\bar{K}(K^\ast)$, $K\bar{K^\ast}(K)$ and $K\bar{K^\ast}(K^\ast)$
intermediate meson loops (see Fig.~\ref{fig-4}). We note that in
Ref.~\cite{achasov-92} the $K\bar{K}(K^\ast)$ loop was also studied
as a major mechanism for $\phi$-$\rho$ mixing.

As follows, we present the transition amplitudes for these meson
loops in $e^+e^-\to \omega\pi^0$. The detailed expressions for the
loop integrals for the $t$-channel transitions have been given in
Ref.~\cite{Li:2007au}. Hence, we only list the necessary formulas
here.

The transition amplitude for $e^+e^- \to\phi\to \omega \pi^0$ via
an intermediate meson loop can be expressed as:
\be
M_{fi}= {\bar v}^{(s^\prime)}(p_e^\prime)(-ie\gamma^\rho)u^s(p_e)
\frac {-ig_{\rho\sigma}} {s} \frac {eM_\phi^2} {f_\phi} \frac
{i\varepsilon_\phi^\sigma} {s-M_\phi^2 + i\Gamma_\phi(s) \sqrt{s}}
\int \frac {d^4p_2}{(2\pi)^4}\sum_{K^* pol}
 \frac {T_1T_2T_3}{a_1a_2a_3}{\cal F}(p_2^2) \ .
 \ee
where the vertex functions for $K\bar{K}(K^*)$ are \be
\left\{\begin{array}{ccl}
 T_1 &\equiv& ig_1(p_1-p_3)\cdot \varepsilon_\phi \\
 \nonumber
 T_2&\equiv& \frac
 {ig_2}{M_\omega}\varepsilon_{\alpha\beta\mu\nu}p_\omega^\alpha\varepsilon_\omega^\beta
 p_2^\mu\varepsilon_2^\nu \\
 T_3&\equiv& ig_3(p_\pi+p_3)\cdot \varepsilon_2\end{array}\right.
 \ee
where $g_1$, $g_2$, and $g_3$ are the coupling constants at the
meson interaction vertices (see Fig. \ref{fig-4}). The four vectors,
$p_\phi$, $p_\omega$, and $p_{\pi^0}$ are the momenta for the
initial $\phi$ and final state $\omega$ and $\pi$ meson; The
four-vector momentum, $p_1$, $p_2$, and $p_3$ are for the
intermediate mesons, respectively, while $a_1=p_1^2-m_1^2,
a_2=p_2^2-m_2^2$, and $a_3=p_3^2-m_3^2$ are the denominators of the
propagators of intermediate mesons.

The vertex functions for the $K\bar{K^*}(K)+c.c.$ loop in
Fig.~\ref{fig-4}(b) are \be \left\{
\begin{array}{ccl}
 T_1 &\equiv &\frac {i f_1}{M_\phi}
 \varepsilon_{\alpha\beta\mu\nu}
 p_\phi^\alpha \varepsilon_\phi^\beta p_3^\mu \varepsilon_3^\nu \ , \nonumber \\
 T_2&\equiv& i f_2(p_1-p_2)\cdot \varepsilon_\omega \ , \nonumber\\
 T_3&\equiv & i f_3(p_\pi-p_2)\cdot \varepsilon_3 \ .
 \end{array}\right.
\ee
where $f_{1,2,3}$ are the coupling constants.

Similarly, we have vertex functions for the intermediate
$K\bar{K^*}(K^*)+c.c.$ loop (Fig.~\ref{fig-4}(c)):
\be\label{vertex-1b} \left\{\begin{array}{ccl}
 T_1 &\equiv &\frac {i h_1}{M_\phi}
 \varepsilon_{\alpha\beta\mu\nu}
 p_\phi^\alpha \varepsilon_\phi^\beta p_3^\mu \varepsilon_3^\nu \ , \\
 \nonumber
 T_2&\equiv& \frac {i h_2}{m_2}
 \varepsilon_{\alpha^\prime\beta^\prime\mu^\prime\nu^\prime}
 p_2^{\alpha^\prime} \varepsilon_2^{\beta^\prime} p_\omega^{\mu^\prime} \varepsilon_\omega^{\nu^\prime} \ , \\
 T_3&\equiv & \frac {i h_3}{m_3}
 \varepsilon_{\alpha^{\prime\prime}\beta^{\prime\prime}\mu^{\prime\prime}\nu^{\prime\prime}}
 p_2^{\alpha^{\prime\prime}} \varepsilon_2^{\beta^{\prime\prime}} p_3^{\mu^{\prime\prime}} \varepsilon_3^{\nu^{\prime\prime}}
\end{array}\right.
 \ee
where $h_{1,2,3}$ are the coupling constants.

In the above three kinds of vertex functions, the coupling constants
are determined via experimental value and SU(3)
relations~\cite{Li:2007au,pdg2006,tornqvist}. The form factor ${\cal
F}(p^2)$, which takes care of the off-shell effects of the exchanged
particles~\cite{Locher:1993cc,Li:1996yn,Li:1996cj}, is usually
parameterized as \be
 {\cal F}(p^2) = \left(\frac {\Lambda^2 - m_{ex}^2}{\Lambda^2 - p^2}\right)^n,
\ee where $n=0, 1, 2$ correspond to different treatments of the loop
integrals. In the present work, we only consider the monopole form
factor, i.e. $n=1$. For the cut-off energy $\Lambda$, it is usually
parameterized:

\bea \Lambda = m_{ex} + \alpha \Lambda_{QCD}, \eea where
$\Lambda_{QCD}=220$ MeV and $\alpha$ is a tunable parameter,
$m_{ex}$ is the mass of exchanged meson. We note that this form of
form factor is slightly different from that adopted in
Ref.~\cite{Li:2007au}. It smooths out the integrals in terms of a
varying $\Lambda$ value.

\subsubsection {$I=0$ transitions via $\phi$-$\rho^0$ and $\omega$-$\rho^0$ strong mixings}

\begin{figure}
\begin{center}
\includegraphics[scale=0.7]{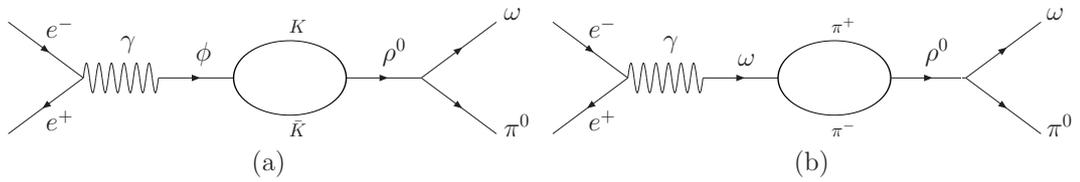}
\caption{The schematic diagrams of $I=0$ transitions for
$e^+e^-\to\omega\pi^0$ via $\phi$-$\rho^0$
mixing.}\protect\label{fig-5}
\end{center}
\end{figure}

The $I=0$ transitions via the $s$-channel $\phi$-$\rho^0$ and
$\omega$-$\rho^0$ strong mixings are shown in Fig.~\ref{fig-5}. The
amplitude for $\phi$-$\rho^0$ mixing via the charged kaon loop can
be expressed as follows:
 \bea  \label{amp-mixing}
 M_{fi}^c &=&{\bar v}(p_{e^+})
(-ie\gamma_a)u(p_{e^-})\frac {g_{ab}(-g^{bc} + \frac {p_\phi^b
p_\phi^c} {M_\phi^2})} {s(s-M_\phi^2+i\Gamma_\phi(s){\sqrt s})}
\frac {eM_\phi^2} {f_\phi}g_{\phi K^+ K^-}g_{\rho K^+ K^-}
\nonumber\\  &\times&  \int\frac {d^4p_2} {(2\pi)^4} \frac
{4p_{2c}p_{2d}} {(p_1^2-m_1^2)(p_2^2-m_2^2)}\frac {-ig^{df}}
{s-M_\rho^2+ i \Gamma(s) {\sqrt s}} \frac {ig_{\omega\rho\pi}}
{M_\omega}
\varepsilon_{efgh} p_\rho^e p_\omega^g \varepsilon_\omega^h \nonumber\\
 &=& -{\bar v}(p_{e^+}) (-ie\gamma^f)u(p_{e^-})  \frac {1}
{s(s-M_\phi^2+i\Gamma_\phi(s){\sqrt s})} \frac {eM_\phi^2}
{f_\phi}g_{\phi K^+ K^-}g_{\rho K^+ K^-} \nonumber\\  &\times& \frac
{|\vec P_1|^3} {6\pi\sqrt {s}}
 \frac {1} {s-M_\rho^2+ i \Gamma(s)
{\sqrt s}} \frac {g_{\omega\rho\pi}} {M_\omega} \varepsilon_{efgh}
p_\pi^e p_\omega^g \varepsilon_\omega^h \ ,
 \eea
where an on-shell approximation following the Cutkosky rule has been
taken and $\vec P_1$ is the three-vector momentum of the on-shell
kaon in the meson loop.

Taking into account the charge-neutral term, the amplitude can be
written as
 \bea
 M_{fi} &=& {\bar v}(p_{e^+}) (-ie\gamma^f)u(p_{e^-})  \frac {1}
{s(s-M_\phi^2+i\Gamma_\phi(s){\sqrt s})} \frac {eM_\phi^2}{f_\phi}
\nonumber\\  &\times & \epsilon_{\phi\rho} \frac {g_{\omega\rho\pi}}
{M_\omega} \varepsilon_{efgh} p_\pi^e p_\omega^g
\varepsilon_\omega^h \ ,
 \eea
where $\epsilon_{\phi\rho}$ is the strong isospin-violating coupling
strength between $\phi$ and $\rho^0$,
 \bea\label{formu-epsilon}
 \epsilon_{\phi\rho}\equiv \frac {1} {6\pi{\sqrt s}D_\rho}
 \left[{g_{\phi K^+K^-}g_{\rho K^+K^-}{\mbox P}_{K^+K^-}^3(s)+g_{\phi K^0{\bar K}^0}g_{\rho
 K^0{\bar K}^0}{\mbox P}_{K^0{\bar K}^0}^3(s)}\right],
 \eea
where $D_\rho\equiv D_\rho(s)=M_\rho^2-s-i\sqrt{s} \Gamma_\rho(s)$
and ${\mbox P}_{K^+K^-}$ and ${\mbox P}_{K^0{\bar K}^0}$ are the
three-vector momentum of the charged and neutral kaons,
respectively. Note that there exists a sign between $g_{\rho
K^+K^-}$ and $g_{\rho K^0{\bar K}^0}$ which brings cancellation
between those two terms on the right-hand side of
Eq.~(\ref{formu-epsilon}). At the mass of the $\phi$ meson, we
obtain $\epsilon_{\phi\rho}=(9.51-i3.31)\times 10^{-4}$ as an
effective strong isospin-violating coupling for $\phi\to\rho^0$.
This result can be compared with the intermediate meson transition
in Eq. (2.10) of Ref.~\cite{achasov-92} apart from factor $D_\rho$,
while the EM part has been contained in Eqs.~(\ref{I0-c}) and
(\ref{I0-d}).

The $\omega$-$\rho^0$ strong mixing occurs only via intermediate
charged pion loop transition as illustrated by Fig.~\ref{fig-5}(b).
The transition amplitude is similar to that for $\phi$-$\rho^0$
mixing, from which the strong isospin-violating coupling strength
can also be defined,
 \bea\label{formu-epsilon-omega}
 \epsilon_{\omega\rho}\equiv \frac {1} {6\pi{\sqrt s}D_\rho}
 g_{\omega \pi^+\pi^-}g_{\rho \pi^+\pi^-}{\mbox
 P}_{\pi^+\pi^-}^3(s).
 \eea
At the mass of $\phi$ meson, we obtain
$\epsilon_{\omega\rho}=(13.7-i4.8)\times 10^{-3}$, which is larger
than that $\epsilon_{\phi\rho}$. However, we note in advance that
the $\omega$-$\rho^0$ mixing effects are negligibly small at the
$\phi$ mass.

 With all the
contributing amplitudes above combined, the total transition
amplitude can be expressed as
 \be\label{phase-delta} M_{fi} = M_{fi}^{I=1}
+ M_{fi}^{I=0},
 \ee
  with
  \be\label{em-loop-phase} M_{fi}^{I=0} =
M_{fi}^{EM-I=0} + e^{i\delta_L}
(M_{fi}^{Loop-I=0}+M_{fi}^{mixing-I=0}),
 \ee
 where $\delta_L$ is the
phase angle for the EM and loop transition amplitudes, and will be
determined by the experimental data. The relative phase between
$M_{fi}^{Loop-I=0}$ and $M_{fi}^{mixing-I=0}$ is given by the
effective Lagrangian.

The differential cross section can thus be obtained:
 \be \frac {d\sigma} {d\Omega} = \frac {1}
{64\pi^2 s} \frac {|{\bf p}_f|} {|{\bf p}_e|} \frac {1}{4}
\sum_{spin}| M_{fi}|^2 \ee
 where ${\bf p}_e$ is the magnitude of the
three-vector momentum of the initial electron (positron) in the
overall c.m. system. We can neglect the mass of the electron, and
have $|{\bf p}_e|=E_{cm}/2$.

In experiment, by defining the background cross section
$\sigma_0(\sqrt{s})$ (from $I=1$ transitions and $I=0$ $\omega$
excitation),  the full cross section can be parameterized
as~\cite{Achasov:1999wr,Achasov:1999jc,Aulchenko}:
 \be\label{parameter-scheme}
 \sigma({\sqrt s})=\sigma_0({\sqrt s})
\left|1-Z\frac {M_\phi\Gamma_\phi} {D_\phi}\right|^2, \ee
 where
$M_\phi$ and $\Gamma_\phi$ are the $\phi$ mass and width and $D_\phi
= M_\phi^2 - s - i{\sqrt{s}} \Gamma_\phi $ is the inverse propagator
of the $\phi$ meson; $Z$ is a complex interference parameter that
equals to the ratio of $\phi$ excitation amplitude to the background
terms and describes the energy evolution of their relative phase. In
our model, apart from calculating the cross sections, we shall also
extract this quantity in order to compare with the values fitted in
experiment~\cite{Ambrosino:2007wf,Ambrosino:2008gb}.

\section{Numerical results}

The $I=1$ components from the low-lying states $\rho$ and
$\rho^\prime(1450)$ play an important role of background to the
$\phi$ meson excitations, while we find that the $\omega$
contribution to the background via $\omega$-$\rho^0$ mixing is
negligibly small in the vicinity of the $\phi$ mass. In the VMD
model the $\rho$ contribution is relatively better controlled with
its well-determined total and partial widths~\cite{pdg2006}. In
contrast, the experimental status for $\rho^\prime(1450)$ is still
not well-established. This unavoidably leads to large uncertainties
with its contributions. Interestingly, it shows that the side-band
cross sections are essential for determining the contributions from
$\rho^\prime$. Since the exclusive contributions from $\rho$ meson
with $M_\rho=775.0$ MeV and $\Gamma_\rho(M_\rho^2)=149.4$ MeV still
significant underestimate the side-band cross sections, it is a
signal indicating that additional $I=1$ components are needed.
Nevertheless, it determines a constructive relative phase between
$\rho$ and $\rho^\prime$ terms in Eq. (\ref{I1-em}), i.e.
$\delta_1=180^\circ$, which is the same
as~\cite{Aulchenko,Clegg:1993mt,Akhmetshin:2003ag}.

By adopting the PDG values for the $\rho^\prime(1450)$, i.e.
$M_{\rho^\prime}=1.459$ GeV,
$\Gamma_{\rho^\prime}(M_{\rho^\prime}^2)=147$ MeV and ${\cal
B}_{\rho^\prime\to\omega\pi^0}=21\%$~\cite{pdg2006}, we find that it
is still far from sufficient to account for the cross sections away
from the $\phi$ mass. In particular, it shows that the cross section
is not sensitive to the $\rho^\prime$ mass, but quite sensitive to
its total width, due to the requirement of a stronger coupling to
$\omega\pi^0$. This suggests that higher $\rho$ states are also
needed. At the $\phi$ mass we find that the effects of including
higher $\rho$ states simply enhance the $I=1$ cross sections and the
behavior of the energy dependence is not sensitive to the presence
of another $\rho$ state, e.g. $\rho(1700)$. Since the total width
for $\rho^\prime(1450)$ still has large uncertainties, we simply
leave it to be determined by the side-band cross sections to take
into account the possible contributions from higher $\rho$ state.

The above features allow us to numerically fit the model parameters
with the new and high-precision data from KLOE. In
Ref.~\cite{Ambrosino:2008gb} the $\phi$ excitation cross sections
are measured in both $e^+ e^-\to \omega\pi^0\to
\pi^+\pi^-\pi^0\pi^0$ and $\gamma\pi^0\pi^0$. With $\omega$
dominantly decaying into $\pi^+\pi^-\pi^0$, much smaller
experimental uncertainties are achieved in the $4\pi$ production
channel than in $\gamma\pi^0\pi^0$. These accurate data provide a
stringent constraint on our model parameters.

In total, we have three parameters to be determined by the
experimental data, i.e. a phase angle $\delta_L$ in
Eq.~(\ref{phase-delta}), the form factor parameter $\alpha$, and the
total width $\Gamma_{\rho^\prime}$ for $\rho^\prime$. Two fits are
carried out as an investigation of the parameter space. In Fit-I, we
fix the phase angle $\delta_L=-90^\circ$ while leave $\alpha$ and
$\Gamma_{\rho^\prime}$ to be determined by the data. In Fit-II, we
free these three parameters to let them be fitted by the data.

In Table~\ref{tab-para}, the fitted parameters are listed. It shows
that with $\delta_L=-90^\circ$ fixed in Fit-I, the reduced $\chi^2$
value is much larger than that in Fit-II. Comparing these two fits,
we see that $\delta_L= -111.6^\circ\pm 2.3^\circ$ is favored by the
data while the fitted values for the other two parameters are
consistent.

We list the branching ratios extracted from these two fits in
Table~\ref{tab-1}, and compare them with the PDG~\cite{pdg2006} and
KLOE results~\cite{Ambrosino:2007wf,Ambrosino:2008gb}. The
$\phi$-$\rho^0$ strong mixing is relatively small and independent of
those parameters. Therefore, it keeps the same for both fits and
contributes an exclusive branching ratio of $0.37\times 10^{-5}$.
The EM amplitudes have dependence on the $\rho^\prime$ width, which
leads to small differences between these two fits. We then notice
that significant differences between these two fits arise from the
meson loop contributions. In Fit-I, the loop contributions
constructively interfere with other transitions and lead to a
relatively larger branching ratio, $BR_{\phi\to\omega\pi^0}= 4.29
\times 10^{-5}$, while in Fit-II the loop transition amplitude has a
destructive effect and gives $BR_{\phi\to\omega\pi^0}= 2.83 \times
10^{-5}$. This is a novel feature arising from the precise data
\cite{Ambrosino:2008gb}.

\begin{figure}
\begin{center}
\includegraphics[scale=0.6]{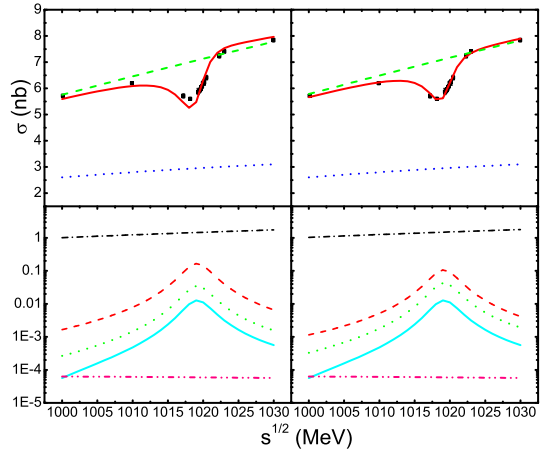}
\caption{ (color online).  The $\sqrt{s}$-dependence of the total
cross section for $e^+ e^-\to \omega\pi^0$ fitted in Fit-I (left)
and Fit-II (right). The data are from KLOE
measurement~\protect\cite{Ambrosino:2008gb}. In the lower panels,
the dotted curves denote the exclusive $t$-channel meson loop
contributions, the solid lines denote the contributions from the
$\phi$-$\rho^0$ mixing, the dot-dot-dashed lines denote the
contributions from the $\omega$-$\rho^0$ mixing, while the dashed
ones for the inclusive cross sections for $\phi$ excitations (EM
plus strong isospin violation). The dot-dashed lines stand for cross
sections from $\rho^\prime(1450)$ with fitted total widths. In the
upper panels, the dotted lines denote the contributions from the
$\rho$ meson, while the dashed lines are for the inclusive
contributions from background including the dominated $I=1$
component ($\rho$ plus $\rho^\prime(1450)$) and a small $I=0$
$\omega$ excitation. The solid curves are the full model results.
}\protect\label{fig-6}
\end{center}
\end{figure}

To disentangle this interesting phenomenon, we plot cross sections
from different transitions in Fig.~\ref{fig-6} for these two fits.
It shows that Fit-II accounts for the data perfectly while Fit-I
exhibits some discrepancies. The following points can be learned:

i) The extracted branching ratio is very sensitive to the line shape
of the $\phi$ meson dip. This can be recognized by comparing the
previous KLOE data~\cite{Ambrosino:2007wf} and the new
one~\cite{Ambrosino:2008gb}. In Ref.~\cite{Ambrosino:2007wf} the
relatively large uncertainties allow us to fix $\delta_L=-90^\circ$
to produce a result similar to Fit-I, i.e. $BR_{\phi\to\omega\pi^0}=
4.29 \times 10^{-5}$. But with data from
Ref.~\cite{Ambrosino:2008gb}, Fit-I exhibit obvious deviations from
the precise data. In contrast, the success of Fit-II with a varying
$\delta_L$ suggests a stringent constraint from the data.

ii) The side-band cross sections are important for determining the
background contributions. As shown by the solid curve for the full
calculation, at the energies away from the $\phi$ mass, the
contributions from the $\phi$ excitation die out quickly, and the
cross sections are dominated by the $I=1$ components.

iii) The extracted branching ratio from Fit-II is smaller than the
result from Refs.~\cite{Ambrosino:2007wf,Ambrosino:2008gb} and
PDG~\cite{pdg2006}. In our calculation it is due to the constraint
from the line shape which requires a destructive phase between the
EM and loop transition amplitudes. We also find that the branching
ratio extraction is very sensitive to the background cross sections.
It should be noted that although we can directly calculate the cross
section for $e^+ e^-\to \omega\pi^0\to 4\pi$, the experimental
extraction of this quantity needs understanding of non-$\omega\pi^0$
background which contribute to the $4\pi$ final
state~\cite{Ambrosino:2008gb}.

iv) Our calculations show that the $\omega$-$\rho^0$ mixing effects
are negligibly small at the $\phi$ mass region as illustrated by the
dot-dot-dashed curved in the lower panels of Fig.~\ref{fig-6}.

\begin{figure}
\begin{center}
\includegraphics[scale=0.6]{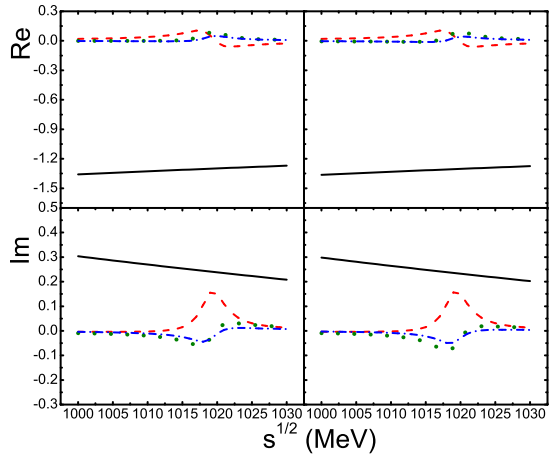}
\vspace{-0.0cm}\caption{(color online). Energy evolution of the real
(upper panel) and imaginary (lower panel) amplitudes given by Fit-I
(left) and Fit-II (right). The solid lines denote the amplitude for
the background terms (dominated by $I=1$ component); the dashed
lines stand for $I=0$ EM transition amplitude, the dotted lines
denote the $I=0$ $t$-channel transitions, while the dash-dot lines
for the $s$-channel $\phi$-$\rho^0$ mixing. A common factor ${\bar
v}^{(s^\prime)}(p_e^\prime)(-ie\gamma^\nu)u^s(p_e)
\varepsilon_{\alpha\beta\mu\nu}{p_\omega^\alpha
\varepsilon_\omega^\beta p_\pi^\mu }$ in the transition amplitudes
has been removed. } \protect\label{fig-7}
\end{center}
\end{figure}

In order to learn more about the underlying dynamics and clarify the
interferences among different transitions, we make a further
analysis of the amplitudes. In Fig.~\ref{fig-7}, we present the real
and imaginary part of the exclusive transitions apart from the
common factor ${\bar v}^{(s^\prime)}(p_e^\prime)(-ie\gamma^\nu)$
$u^s(p_e) \varepsilon_{\alpha\beta\mu\nu}{p_\omega^\alpha
\varepsilon_\omega^\beta p_\pi^\mu }$. The solid lines denote the
real and imaginary part of the background amplitude while the others
are $\phi$ excitation amplitudes from EM, $t$-channel loop
transitions, and $s$-channel $\phi$-$\rho^0$ mixing. Their relative
phases respect to the background terms are included. The visible
difference between Fit-I (left panel) and Fit-II (right panel) is
rather small though it can be seen that the $t$-channel loop
amplitude plays a role to change the line shape due to the change of
phase angle $\delta_L$.

\begin{figure}
\begin{center}
\includegraphics[scale=0.6]{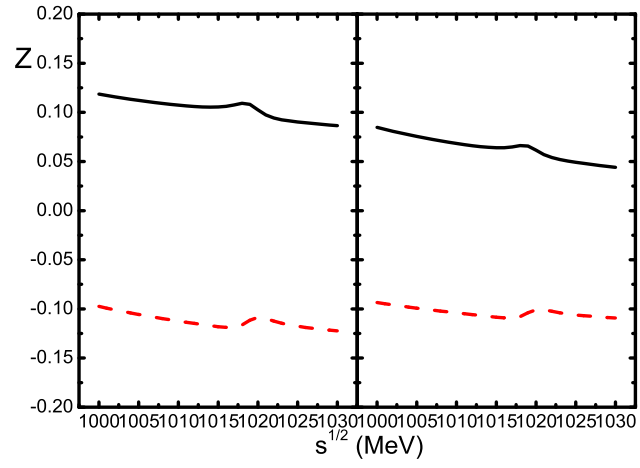}
\vspace{-0.0cm}\caption{(color online).  Energy evolution of
parameter $Z$ extracted from Fit-I (left) and Fit-II (right). The
solid lines stand for the real part while the dashed lines for the
imaginary part.  }\protect\label{fig-8}
\end{center}
\end{figure}

By extracting the $Z$ parameter defined in
Eq.~(\ref{parameter-scheme}), such a small difference can be
highlighted due to the change of the background in these two fits.
As shown by Fig.~\ref{fig-8}, both the real and imaginary part of
$Z$ turn out to be flat in terms of $\sqrt{s}$. The slight
energy-dependence shows that the background contributions (dominated
by $I=1$ component) cannot come from a single isovector excitation.
Their signs clearly indicate the destructive interferences between
the $\phi$ excitation and the background terms.

At the $\phi$ mass, we extract the real and imaginary part of $Z$
and compare them with the experimental results from
KLOE~\cite{Ambrosino:2007wf,Ambrosino:2008gb} etc. As shown by
Tab.~\ref{tab-2}, the extracted real and imaginary part of $Z$ at
the $\phi$ mass are basically in agreement with the
data~\cite{Ambrosino:2007wf,Achasov:1999wr,Achasov:1999jc,Aulchenko}
except that $\mbox{Re} Z$ from Fit-II appears to be relatively
small. This is again due to the difference of the background terms
denoted by the extracted values for $\sigma_0^{4\pi}(M_\phi)$. It
can be shown that with $\sigma_0^{4\pi}(M_\phi)=7.12 $ nb,
$\mbox{Re} Z=0.065\pm 0.002$, and $\mbox{Im} Z=-0.103\pm 0.005$, the
cross section for $e^+ e^-\to \omega\pi^0\to 4\pi$ at
$\sqrt{s}=M_\phi$ can be reproduced. Comparing our Fit-II results
with Ref.~\cite{Ambrosino:2008gb}, the sensitivity of the $\phi$
branching ratio to the background terms can be recognized. Note that
the values for $\sigma_0^{4\pi}(M_\phi)$ in Tab.~\ref{tab-2} from
Refs.~\cite{Ambrosino:2007wf,Ambrosino:2008gb} include about $12\%$
contaminations from resonance background.

\begin{figure}
\begin{center}
\includegraphics[scale=0.8]{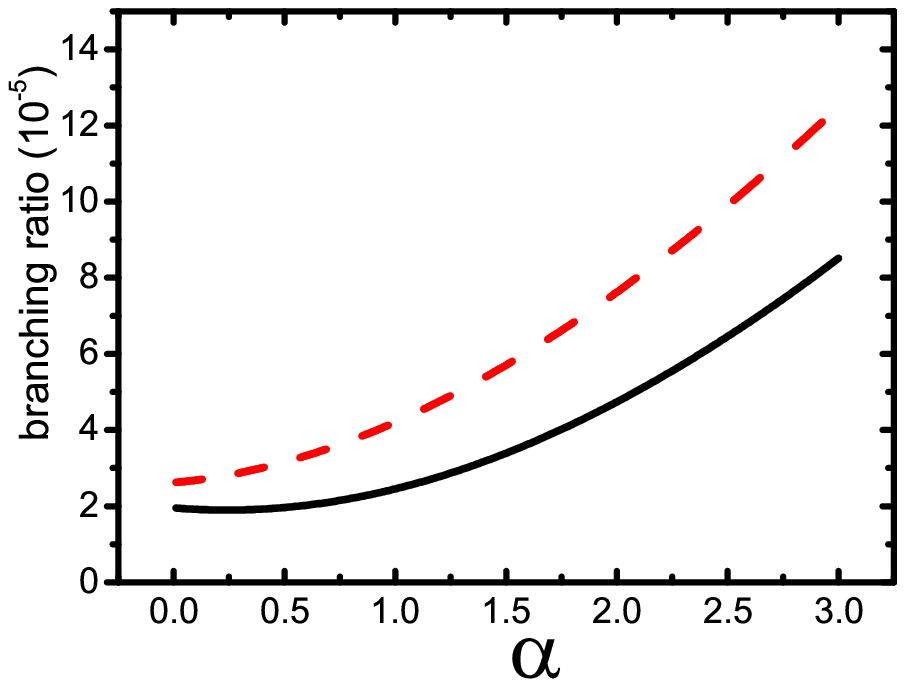}
\vspace{-0.0cm}\caption{ The $\alpha$-dependence of the
$\phi\to\omega\pi^0$ branching ratio in Fit-I (dashed) and Fit-II
(solid). }\protect\label{fig-9}
\end{center}
\end{figure}

In Fig.~\ref{fig-9}, we plot the branching ratio dependence on the
form factor parameter $\alpha$ with $\delta_L$ and
$\Gamma_{\rho^\prime}$ fixed as in Fit-I and II. It shows that over
a broad range of $\alpha$, the branching ratio does not vary
drastically. The uncertainties of the fitted values for $\alpha$ in
Fit-I and II are about $5\%$ (see Tab.~\ref{tab-para}), which
suggest a well constraint of this parameter. It is also an
indication that the loop integrals are well tamed.

\section{Summary}

The precise measurement of the $\phi$ excitation in
$e^+e^-\to\omega\pi^0$ \cite{Ambrosino:2008gb} as a consequence of
interferences between the $I=1$ and $I=0$ components in the
transition amplitudes provides an opportunity to study the isospin
violation mechanisms at low energies. We investigate this process by
quantifying the isospin violations in both EM and strong
transitions. The $I=1$ and $I=0$ EM contributions are studied in the
VMD model, while the $I=0$ strong isospin violating process is
described by the $t$-channel OZI-rule-evading
intermediate-mesons-exchange loops and $s$-channel $\phi$-$\rho^0$
mixing. This is an improved effective Lagrangian approach compared
with our early study in Ref.~\cite{Li:2007au} where the $s$-channel
$\phi$-$\rho^0$ mixing was overlooked.

In $e^+ e^-\to\omega\pi^0$, the side-band cross sections provide a
strong constraint on the $I=1$ EM transitions. Contributions from
$\omega$ meson excitation is also included though it turns out to be
negligibly small. In the calculation we find that the cross section
of $e^+e^-\to\omega\pi^0$ has evident width-dependence on the
$\rho^\prime$ which contributes to the $I=1$ transition amplitudes.
In the previous work for $\phi\to\omega\pi^0$~\cite{Li:2007au}, we
did not include the $\rho^\prime$ due to lack of constraint on it.
By fitting the new KLOE data~\cite{Ambrosino:2008gb}, we succeed in
constraining the model parameters, i.e. phase angle $\delta_L$, form
factor parameter $\alpha$, and the total width
$\Gamma_{\rho^\prime}$ of $\rho^\prime$. We show that there exists a
strong correlation between the OZI-rule-evading and strong isospin
violation mechanisms. We also find that the extracted branching
ratio for $\phi\to\omega\pi^0$ is very sensitive to the line shape
of the cross sections in the vicinity of $\phi$ excitation and the
background estimate. Our result turns to be smaller than that given
by Ref.~\cite{Ambrosino:2008gb}.

It is worth noting that a relatively broad $\rho^\prime$ state is
strongly favored by the newly published KLOE
data~\cite{Ambrosino:2007wf}. Although it is not possible to
answer whether an additional broad $\rho^\prime$ in addition to
the $\rho^\prime(1450)$ is needed, our results shed some light on
this puzzling question in its interference with isospin violating
transitions. It is also worth mentioning that there exist some
hints of broad $\rho^\prime$ states around 1.2$\sim$ 1.6 GeV from
experiments~\cite{crystal-barrel,bes-2006}. Further theoretical
investigation on $e^+e^-\to \omega\pi^0$ over a broader energy
region, and precise measurement of the cross sections may be able
to clarify such questions.

\section*{Acknowledgement}

Useful discussions with C. Hanhart, S. Narision, J.-M. Richard, and
B.S. Zou are acknowledged. This work is supported, in part, by the
National Natural Science Foundation of China (Grants No. 10675131
and 10491306), Chinese Academy of Sciences (KJCX3-SYW-N2), and the
U.K. EPSRC (Grant No. GR/S99433/01).

\begin{table}[ht]
\begin{tabular}{|c|c|c|c|c|c|c|c|} \hline
         Parameter  &$\theta$         &$\alpha$    & $\Gamma_{\rho^\prime}$(MeV)  & $\chi^2/d.o.f$\\ [1ex] \hline
         Fit-I      &$-90.0^\circ$(fixed)          &$1.125\pm 0.052$         &$674\pm 6$    & 5.71 \\ [1ex] \hline
         Fit-II     &$-111.6^\circ\pm 2.3^\circ$   &$1.244\pm 0.051$         &$683\pm 6$    & 0.54 \\ [1ex] \hline
\end{tabular}
\caption{ The parameters fitted in Fit-I and Fit-II schemes. }
\label{tab-para}
\end{table}

\begin{table}[ht]
\begin{tabular}{|c|c|c|c|c|c|c|c|ccc} \hline
         Branching ratio ($\times 10^{-5}$) &EM         &Meson loop    & $\phi$-$\rho^0$ mixing &Total     &PDG~\cite{pdg2006}          &Exp~\cite{Ambrosino:2007wf} & Exp~\cite{Ambrosino:2008gb}   \\ [1ex] \hline
          \parbox[c]{1cm} {Fit-I  \\  Fit-II}
         & \parbox[c]{1cm} {2.95  \\ 2.97   }
         & \parbox[c]{1cm} {  0.93 \\ 1.14 }
         & \parbox[c]{1cm}  {0.37  \\ 0.37}
         & \parbox[c]{1cm}  { 4.29  \\ 2.83}
         & $5.2^{+1.3}_{-1.1}$
         & $5.63\pm  0.70$
         & $4.4\pm 0.6$   \\ [1ex] \hline
\end{tabular}
\caption{ Branching ratios for $\phi\to \omega\pi^0$ extracted from
our model with two different fitting schemes. Experimental
data~\cite{pdg2006,Ambrosino:2007wf,Ambrosino:2008gb} and exclusive
branching ratios from EM, $t$-channel meson loop transitions, and
$s$-channel $\phi$-$\rho^0$ mixing are also listed. } \label{tab-1}
\end{table}

\begin{table}[ht]
\begin{tabular}{|c|c|c|c|c|c|c|c|} \hline
                                            &$\sigma_0^{4\pi}(M_\phi),(nb)$  &Re Z                                &Im Z                          \\ [1ex] \hline
         Ref.~\protect\cite{Achasov:1999wr}         &$7.32\pm 0.14 \pm 0.38$         &$0.110\pm 0.019 \pm 0.003$          &$-0.129\pm 0.025 \pm 0.005$   \\ [1ex] \hline
         Ref.~\protect\cite{Achasov:1999jc}         &$7.28\pm 0.18 \pm 0.80$         &$0.104\pm 0.028 \pm 0.006$          &$-0.118\pm 0.030 \pm 0.009$   \\ [1ex] \hline
         Ref.~\protect\cite{Aulchenko}              &$7.34\pm 0.14$                  &$0.112\pm 0.019$                    &$-0.129\pm 0.025$             \\ [1ex] \hline
         Ref.~\protect\cite{Ambrosino:2007wf}       &$(8.12\pm 0.14)^\dag$                  &$0.097\pm 0.012$                    &$-0.133\pm 0.009$             \\ [1ex] \hline
         Ref.~\protect\cite{Ambrosino:2008gb}       &$(7.89\pm 0.06\pm 0.07)^\dag$          &$0.106\pm 0.007\pm 0.004$           &$-0.103\pm 0.004\pm 0.003$    \\ [1ex] \hline
         Fit-I                              &$7.08\pm 0.04$                  &$0.107\pm 0.005$                    &$-0.110\pm 0.009$             \\ [1ex] \hline
         Fit-II                             &$7.12\pm 0.03$                  &$0.065\pm 0.002$                    &$-0.103\pm 0.005$             \\ [1ex] \hline

\end{tabular}
\caption{ The interference parameter $Z$ extracted from our model by
two different fitting schemes at the $\phi$ mass is compared with
experimental results. Note that the values of
$\sigma_0^{4\pi}(M_\phi)$ labeled by ``$\dag$" contain about $12\%$
non-$\omega\pi^0$
contaminations~\protect\cite{Ambrosino:2007wf,Ambrosino:2008gb}. }
\label{tab-2}
\end{table}


\begin{thebibliography}{99}
\bibitem{Ambrosino:2007wf}
  F.~Ambrosino {\it et al.}  [KLOE collaboration],
  arXiv:0707.4130 [hep-ex].
%

\bibitem{Ambrosino:2008gb}
  F.~Ambrosino {\it et al.}  [KLOE collaboration],
  Phys.\ Lett.\  B {\bf 669}, 223 (2008)
  [arXiv:0807.4909 [hep-ex]].



\bibitem{Li:2007au}
  G.~Li, Q.~Zhao and B.~S.~Zou,
  Phys.\ Rev.\  D {\bf 77}, 014010 (2008)
  [arXiv:0706.0384 [hep-ph]].
%
%
\bibitem{pdg2006}
W. M. Yao {\it et. al.} [Particle Data Group], J. Phys. G {\bf 33},
1 (2006).
%
\bibitem{Miller:1990iz}
  G.~A.~Miller, B.~M.~K.~Nefkens and I.~Slaus,
  Phys.\ Rept.\  {\bf 194}, 1 (1990).

%
\bibitem{bramon-81} A. Bramon, Phys. Rev. D {\bf 24}, 1994 (1981).
%
\bibitem{donoghue} J.F. Donoghue, B.R. Holstein, and D. Wyler,
Phys. Rev. Lett. {\bf 69}, 3444 (1992).
%
\bibitem{ametller} L.~Ametller, C.~Ayala and A.~Bramon,
  Phys.\ Rev.\  D {\bf 30}, 674 (1984).

%
\bibitem{coon-86} S.A. Coon, B.H.J. McKellar, and M.D. Scadron,
Phys. Rev. D {\bf 34}, 2784 (1986).
%
\bibitem{coon-87} S.~A.~Coon and R.~C.~Barrett,
  Phys.\ Rev.\  C {\bf 36} (1987) 2189.

%
\bibitem{genz} H. Genz and S. Tatur, Phys. Rev. D {\bf 50}, 3263
(1994).
%
%
\bibitem{Bauer:1975bw}
  T.~Bauer and D.~R.~Yennie,
  Phys.\ Lett.\  B {\bf 60}, 169 (1976).
%
\bibitem{Akhmetshin:2003ag}
  R.~R.~Akhmetshin {\it et al.}  [CMD-2 Collaboration],
  Phys.\ Lett.\  B {\bf 562}, 173 (2003)
  [arXiv:hep-ex/0304009].
%
\bibitem{Eletsky:1982py}
  V.~L.~Eletsky, B.~L.~Ioffe and Y.~I.~Kogan,
  Phys.\ Lett.\  B {\bf 122}, 423 (1983).
%
\bibitem{Narison:1982yc}
  S.~Narison and N.~Paver,
  Z.\ Phys.\  C {\bf 22}, 69 (1984).
%
\bibitem{Lublinsky:1996yf}
  M.~Lublinsky,
  Phys.\ Rev.\  D {\bf 55}, 249 (1997)
  [arXiv:hep-ph/9608331].

\bibitem{Li:2007ky}
  G.~Li, Q.~Zhao and C.~H.~Chang, J. Phys. G: Nucl. Part. Phys. {\bf 35}, 055002 (2008).
  arXiv:hep-ph/0701020.

\bibitem{kloe-phi-rho-pi} A. Aloisio {\it et al.} [KLOE Collaboration], Phys. Lett. B {\bf 561},
55 (2003); {\bf 609}, 449(E) (2005); arXiv:hep-ex/0303016.
%
\bibitem{Lipkin:1986bi}
  H.~J.~Lipkin,
  Nucl.\ Phys.\  B {\bf 291}, 720 (1987).
%
\bibitem{Lipkin:1986av}
  H.~J.~Lipkin,
  Phys.\ Lett.\  B {\bf 179}, 278 (1986).
%
\bibitem{achasov-92} N.N. Achasov and A.A. Kozhevnikov,
  Int. J. Mod. Phys. {\bf A 7}, 4825 (1992).
%
\bibitem{tornqvist} N.A. Tornqvist,
  Annals Phys.\  {\bf 123}, 1 (1979).
\bibitem{Locher:1993cc}
  M.~P.~Locher, Y.~Lu and B.~S.~Zou,
  Z.\ Phys.\  A {\bf 347}, 281 (1994)
  [arXiv:nucl-th/9311021].
%
\bibitem{Li:1996yn}
  X.~Q.~Li, D.~V.~Bugg and B.~S.~Zou,
  Phys.\ Rev.\  D {\bf 55}, 1421 (1997).
\bibitem{Li:1996cj}
  X.~Q.~Li and B.~S.~Zou,
  Phys.\ Lett.\  B {\bf 399}, 297 (1997)
  [arXiv:hep-ph/9611223].

\bibitem{Achasov:1999wr}
  M.~N.~Achasov {\it et al.},
  Nucl.\ Phys.\  B {\bf 569}, 158 (2000)
  [arXiv:hep-ex/9907026].
\bibitem{Achasov:1999jc}
  M.~N.~Achasov {\it et al.},
  Phys.\ Lett.\  B {\bf 449}, 122 (1999)
  [arXiv:hep-ex/9901020].
\bibitem{Aulchenko}
  V.~M.~Aulchenko {\it et al.},
  Jou.\ ExP.\ Th.\ Phys. {\bf 90}, 927 (2000).
%
\bibitem{Clegg:1993mt}
  A.~B.~Clegg and A.~Donnachie,
  Z.\ Phys.\  C {\bf 62}, 455 (1994).




\bibitem{crystal-barrel} C. Amsler {\it et al.} [Crystal Barrel Collaboration], Nucl. Phys.
{\bf A 740}, 130 (2004).

\bibitem{bes-2006} M. Ablikim {\it et al.} [BES Collaboration],
Phys. Rev. Lett. {\bf 97}, 142002 (2006).



\end{thebibliography}
\end{document}